\documentclass[ aps,prb,amsmath,amssymb,preprint,floatfix]{revtex4-1}
\usepackage{graphicx}
\usepackage{bm}

\begin{document}
\title[]{Finite element solution of the Fokker-Planck equation for single domain particles}

\author{N. V. Peskov}
\affiliation{Faculty of Computational Mathematics and Cybernetics, \\Lomonosov Moscow State University,  Moscow, Russian Federation.}

\date{\today}

\begin{abstract}

The Fokker-Planck equation derived by Brown for the probability density function of the orientation of the magnetic moment of single domain particles is one of the basic equations in the theory of superparamagnetism. Usually this equation is solved by expanding the solution into a series of spherical harmonics, which in this case is a complex and cumbersome procedure. This article presents the implementation procedure and some results of the numerical solution of the Fokker-Planck equation using the finite element method. A method for creating a sequence of triangular grids on the surface of a sphere based on an inscribed icosahedron is described. The equations of the finite element method are derived and examples of numerical solutions are presented. The processes of magnetization and demagnetization under heating of a particle with cubic magnetic anisotropy are simulated.

 \end{abstract}
\maketitle

\section{Introduction}

A particle of ferromagnetic material below a certain critical size (typically 30 nm in diameter) constitutes a single domain particle meaning that it is in a state of uniform magnetization for any applied field \cite{1}. The magnetic moment $\bm M$ per unit volume of such a particle can be represented as a vector of constant magnitude, $\bm M=M_s\bm u$, $|\bm u|=1$, where $M_s$ is the saturation magnetization per unit volume. In the course of time only the orientation of the magnetic moment, determined by the unit vector $\bm u$, can change. The orientation is influenced by internal magnetocrystalline anisotropy, an external magnetic field, and random fluctuations caused by thermal agitation. 

The magnetic properties of single domain particles in the absence of thermal fluctuations described by Stoner and Wohlfarth \cite{2}. The theory of the thermal fluctuations of the magnetization of single domain particles was began with work of N\'eel \cite{3} and was further developed by Brown \cite{4}. 

A particle with orientation $\bm u = (\sin\theta\cos\phi,\sin\theta\sin\phi,\cos\theta)$ in a Cartesian coordinate system, where $\theta$ is the polar angle and $\phi$ is the azimuth, is assumed to be in internal thermodynamic equilibrium at temperature $T$, with Helmholtz free energy per unit volume $E_a(\theta,\phi,T)$. The particle is not necessarily in external equilibrium with the applied field $\bm H=H\bm h$, $|\bm h|=1$. The Gibbs free energy per unit volume is $V(\theta,\phi,T,\bm H) = E_a(\theta,\phi,T)-M_sH(\bm u\cdot\bm h)$, which will be written below as $V(\theta,\phi)$.

In the absence of thermal agitation, changes of $\bm u$ are assumed to obey the Gilbert equation \cite{5}
\begin{equation}
\frac{d\bm u}{dt}= -\frac{h^\prime M_s}{\alpha}\left(\bm u\times
\nabla V\right) + h^\prime\left(\bm u\times\left(\bm u\times
\nabla V\right)\right),
\label{ge}
\end{equation}
where $t$ is the time, $\alpha$ is a dimensionless damping coefficient and
$h^\prime = (\alpha\gamma)/((1+\alpha^2)M_s)$, $\gamma$ is the ratio of magnetic moment to angular momentum, and $\nabla$ is the angular part of the gradient.

In Ref.~4 the evolution of the magnetic moment was considered as a Brownian motion along the surface of a unit sphere of a point,  representing the orientation of the magnetic moment, subjected to the applied field and magnetic anisotropy. As the Langevin equation for this motion Brown took Gilbert's equation, supplemented by a random Gaussian white noise field, which takes into account the collisional damping. Using the obtained Langevin equation, Brown derived the Fokker-Planck equation (FPE) for the probability density function $W(\theta,\phi,t)$ of orientations of magnetic moments, i.e. representing points on the unit sphere. 

The FPE derived by Brown can be written in the form of a continuity equation
\begin{equation}
\label{b0}
\frac{\partial W}{\partial t} - \nabla\cdot \left(k\nabla W + \frac{d\bm u}{dt}W\right)=0,
\end{equation}
where the coefficient $k$ should be chosen so that the Boltzmann distribution $W_B\propto \exp(-(vV)/(k_BT))$ is a stationary solution of (\ref{b0}) for a particle of volume $v$, $k_B$ is the Boltzmann constant.

Brown's approach to the theory of magnetism of single domain particles opened up the possibility of applying the methods developed in the theory of Brownian motion to the study of a superparamagnetism. A comprehensive review of these methods, as well as the most important results obtained with their help, are presented in the book \cite{6} by Coffey and Kalmykov. Since the present paper is devoted to the numerical solution of Brown's FPE, we restrict ourselves to a brief overview of commonly used methods for solving this equation.

The solution of the equation (\ref{b0}) is usually associated with a decomposition of $W$ in a basis consisting of angular eigenfunctions of the Laplace operator (spherical harmonics), which results in an infinite system of differential-recurrent equations for the coefficients of decomposition. The procedure for deriving of this system from the FPE equation is given in Ref.~7. The system of differential-recurrent equations has the form
\begin{equation}
\label{x}
\frac{d}{dt}\bm X(t)=\mathbf A \bm X(t),
\end{equation}
where $\bm X=\{x_0(t),x_1(t),\dots\}$ is the infinite vector of expansion  coefficients, and $\mathbf A$ is a matrix that can depend on time. 

One of the peculiarities of the system (\ref{x}) is that if the elements of the vector $\bm X$ are properly ordered, then the matrix $\mathbf A$ becomes the $d$-diagonal matrix for any magnetic anisotropy that can be expressed as the finite combination of spherical harmonics. The number of diagonals, $d$, is determined by the type of magnetic anisotropy. For isotropic particles, $d = 3$, for anisotropic particles, $d> 3$.

The time dependence of the matrix $\mathbf A$ may, in particular, arise due to the time dependence of the applied field $\bm H$. In studies related to the simulation of dynamic magnetic hysteresis or the calculation of dynamic magnetic susceptibilities a periodic applied field (ac field), $\bm H(t)=\bm H_0\cos\omega t$ is usually considered. This field generates a time dependence through $\cos\omega t$ of some elements of $\mathbf A$. It seems that, for the first time, a study of dynamic hysteresis using the numerical solution of FPE, transformed in the form (\ref{x}), was undertaken in Ref.~8, where the hysteresis of isotropic superparamagnets was studied.

For anisotropic particles for solving FPE under ac field the system (\ref{x}) is reduced to a linear algebraic system for the coefficients $F_m^n$ by substituting
\[x_m(t) = \sum_{n=-\infty}^\infty {F_m^n e^{\imath n\omega t}},\;m=0,1,\dots.\]
The obtained linear system can be solved with the matrix sweep algorithm \cite{9}, or by the matrix continued fraction method \cite{10}.

In connection with linear systems, continued fractions appear, in particular, when solving a linear system with a 3-diagonal matrix by successively eliminating unknowns (Gauss method). For anisotropic particles, the equation system for the coefficients $F_m^n$ has a $d$-diagonal matrix, $d>3$. However, by grouping the unknowns into sub-vectors of the same length so that for sub-vectors it is possible to obtain a linear system with a 3-diagonal matrix, whose elements will be matrices of small dimension \cite{10}. Therefore the solution of the new system can be obtained as a matrix-valued continued fraction \cite{11}. 

The process of magnetic relaxation of single-domain particles can be described by the FPE with a constant applied field $\bm H = \bm H_0$ (dc field). For a dc field, the system (\ref{x}) can also be reduced to a linear algebraic system using the Laplace transform of $X(t)$ and solved by the matrix continued fraction method. Over the past two decades, many physical problems associated with single domain particles have been solved using the matrix continued fraction method. Various examples of such problems can be found in Ref.~6.

An alternative method for investigating the statistical properties of single-domain particles is the Monte Carlo method \cite{12}, which allows one to obtain macroscopic observables by averaging microscopic ones. One of the difficulties in applying the Monte Carlo method to the study of superparamagnetism is the uncertainty of the time scale of the Monte Carlo steps. Using a numerical FPE solution is expected to help overcome this difficulty \cite{13,14}.

This work continues and develops the theme begun in the previous paper \cite{15} and devoted to the application of the finite element method (FEM) for solving Brown's FPE. The FEM approach to FPE is relatively simple and independent of the type of magnetic anisotropy. FEM directly gives the FPE solution, not spherical harmonics, as the matrix continued fractions gives. The probability density function provided by FEM solution of FPE enables one to calculate any statistical characteristics of the single domain particle magnetization.

This work uses the deterministic procedure for creating a triangular grid on the surface of a sphere based on an inscribed icosahedron, which is much simpler and more efficient than the random number procedure described in the previous article \cite{15}. In Ref.~15, a FPE solution was demonstrated with an applied ac field simulating dynamic magnetic hysteresis. Here are the solutions of the PFE with a dc field simulating the magnetization of a particle and its demagnetization with increasing temperature. The examples are calculated for cubic magnetic anisotropy taking into account two anisotropy constants.

\section{Finite element scheme}

Substituting $d\bm u/dt$ from Eq. (\ref{ge}) into Eq. (\ref{b0}), after some transformations, equation (\ref{b0}) can be written in a form convenient for applying the finite element method:
\begin{equation}
\label{b2}
\frac{\partial W}{\partial \tau} = \nabla\cdot\left[\nabla W + W\left(\nabla\tilde V + \frac{1}{\alpha} \left(\bm u\times\nabla\tilde V\right)\right)\right], 
\end{equation}
where $\tilde V = vV/k_BT$; $\tau = t/2\tau_N$ is the dimensionless time,
\[\tau_N = \frac{vM_s(1+\alpha^2)}{2k_BT\gamma\alpha}\]
is the characteristic relaxation time.

The next step of construction the FEM scheme is the generation of a triangular grid on the surface of the sphere.

\subsection{Triangular grid}

In the present paper, for the finite element method, a regular triangular grid is constructed on the surface of the sphere. Here, regularity means that the positions of the grid nodes are calculated using a deterministic procedure and are not random, as was the case in the previous work \cite{15}.

An easy way to cover the sphere with triangles is to build a uniform triangular grid on the surface of the inscribed icosahedron and transfer it to the surface of the sphere using the central projection. Such a grid can be called a 'raw grid'\cite{16}. The faces of the icosahedron are equal equilateral triangles. Therefore, it is possible to build a uniform grid on the surface of the icosahedron by dividing each of its edges into $n$ equal segments and connecting the dividing points lying on adjacent edges with lines parallel to the closing edge. The result is a uniform grid consisting of $N_t=20n^2$ triangles and $N_p=10n^2+2$ nodes. When one transfers the grid to the surface of the sphere using the central projection, the initially uniform grid will be distorted. The degree of grid distortion is usually characterized by two parameters: the ratio $r_1$ of the lengths of the shortest and the longest linear elements, and the ratio $r_2$ of the areas of the smallest and largest grid cells. For the 'raw grid' with $n=81$ (presented below results were obtained with this $n$) one has $r_1=0.686$, and $r_2=0.349$.

Sophisticated methods of optimizing the “raw grid” were developed to reduce its distortion, that is, increase the values of $r_1$ and $r_2$. In particular, the authors of Ref.~16 report an optimized grid with $r_1=0.786$ and $r_2=0.952$ for the number of nodes close to $N_p$ at $n=81$. Since the main purpose of this paper is to demonstrate the potential of FEM to apply to the Brown equation, a very perfect grid that is difficult to build is not necessary here. Therefore, the grid used in the further calculations is constructed in the following rather simple way.

\begin{figure}[h]
\centering
\includegraphics{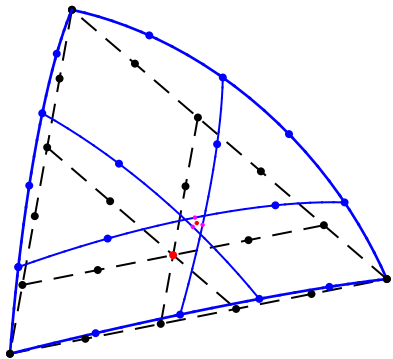}
\caption{Mapping a triangular grid from the icosahedron face to the surface of a sphere}
\label{fig1}
\end{figure}

The basic grid is the uniform grid on the surface of the icosahedron described above. Instead of the central projection, the following procedure is used to transfer the grid to the sphere. The vertices of the inscribed icosahedron remain in their places. The edges of the icosahedron, shown in Figure 1 by thick dashed lines, are mapped into arcs of large circles connecting adjacent vertices, which are shown by blue lines. Grid points on an arc, similar to grid points on an edge, divide each arc into $n$ equal segments. 

Each node of the base grid on the face of the icosahedron lies at the intersection of three straight lines parallel to the edges that limit this face and passing through certain opposite nodes on the icosahedron edges. One of such nodes is shown in Fig.~1 by red dot. The analogues of these lines on the sphere are arcs of three great circles passing through related nodes lying on arcs corresponding to edges. Arcs intersect in pairs, but all three do not intersect at one point. Therefore, the center of a spherical triangle with vertices located at the points of pairwise intersection of the arcs is taken as the image on the surface of the sphere of the base grid node. This node is also indicated by red dot in Fig.~1. The grid constructed in this way for $n = 81$ has the following distortion parameter values: $r_1 = 0.851$, $r_2 = 0.898$.

\begin{figure}[h]
\centering
\includegraphics{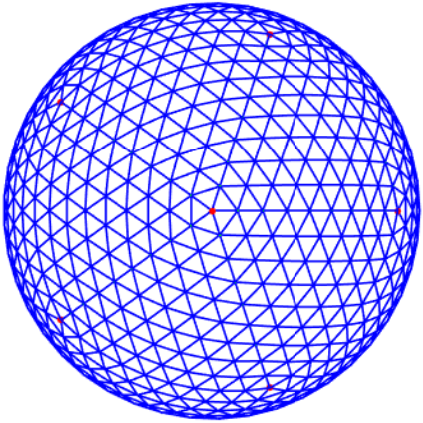}
\caption{Triangular grid with $n = 9$, $N_p=812$, $N_t=1620$. Red dots indicate the vertices of the inscribed icosahedron.}
\label{fig2}
\end{figure}

As an illustration of the grid constructed in this way, a triangular grid on a sphere at $n = 9$ is shown in Fig.~2. Twelve nodes coinciding with the vertices of the inscribed icosahedron (marked in red in Fig. 2) each have 5 nearest neighbors. All other nodes have 6 nearest neighbors each.

\subsection{Finite element equations}

The triangular grid constructed above can be considered as a polyhedron inscribed in the sphere with flat triangular faces and vertices $P_i$, $i = 1,2,\dots,N_p$, which are grid nodes. Let $O_i$ be a neighborhood of the node $P_i$, i.e. the union of triangular faces in which $P_i$ is one of the vertices. A neighborhood $O_i$ is composed of five adjacent triangles with a common vertex $P_i$, if this vertex is the vertex of the inscribed icosahedron, or of six adjacent triangles, if their common vertex is not the vertex of the icosahedron. On the surface of the polyhedron we define a real continuous function $\varphi_i$, $0\leq\varphi_i\leq1$, so that it is linear on each triangle, $\varphi_i(P_i)=1$ and $\varphi_i\equiv 0$ outside of $O_i$. Functions $\varphi_i$, $i = 1,2,\dots,N_p$ will be called finite elements.

Since there is a one-to-one mapping (central projection) between the polyhedron and the sphere, any function defined on the polyhedron can be considered on the sphere, and vice versa. Therefore, both sides of Eq.~(\ref{b2}) can be multiplied by function $\varphi_i$ and integrated over the surface of the sphere
\[ 
\int{\varphi_i\frac{\partial W}{\partial \tau}\,d\Omega} =
\int{\varphi_i\nabla\cdot\left[\nabla W + W\left(\nabla\tilde V + \frac{1}{\alpha} \left(\bm u\times\nabla\tilde V\right)\right)\right]\,d\Omega}.
\] 
Now we apply the Green formula to the integral on the right-hand side, in which there will be no integral over the boundary, since the surface of the sphere has no boundary.
\begin{equation}
\label{f0}
\int{\varphi_i\frac{\partial W}{\partial \tau}\,d\Omega} =
-\int{\nabla\varphi_i\cdot\left[\nabla W + W\left(\nabla\tilde V + \frac{1}{\alpha} \left(\bm u\times\nabla\tilde V\right)\right)\right]\,d\Omega}.
\end{equation} 
The solution to the last equation will be sought in the form\begin{equation}
\label{w}
W(\theta,\phi,\tau)=\sum_{j=1}^{N_p}{W_j(\tau)\,\varphi_j(\theta,\phi)}.
\end{equation}

Substituting (\ref{w}) into (\ref{f0}) one obtains the following system of linear ordinary differential equations
\begin{equation}
\label{f1}
\mathbf M\dot{\bm W} = -\left(\mathbf L + \mathbf F + \alpha^{-1}\mathbf G \right)\bm W,
\end{equation}
where $\bm W$ is the vector $(W_1,W_2,\dots,W_{N_p})^T$, dot denotes the derivative with respect to $\tau$. $\mathbf M$, $\mathbf L$, $\mathbf F$ and $\mathbf G$ are square matrices of dimension $N_p\times N_p$. Matrix elements are calculated by the following formulas
\begin{eqnarray}
&&m_{ij}=\int{\varphi_i\varphi_j\,d\Omega}, \nonumber \\
&& l_{ij}=\int{(\nabla\varphi_i\cdot\nabla\varphi_j)\,d\Omega}, \\
&& f_{ij}=\int{(\nabla\varphi_i\cdot\nabla \tilde V)\,\varphi_j\,d\Omega},\nonumber \\
&&g_{ij}=\int{(\nabla\varphi_i\cdot(\bm u\times\nabla\tilde V))\,\varphi_j\,d\Omega}. \nonumber
\end{eqnarray}

To calculate the matrix elements, the integrals over the sphere are approximated by the integrals over the surface of the embedded polyhedron corresponding to the triangular grid. And by virtue of the definition of finite elements $\varphi_i$, the integration domain is reduced to the intersection of neighborhoods $O_{ij}=O_i\cap O_j$. The set $O_{ij}$ at $i\neq j$  consists of two adjacent triangles with a common side $P_iP_j$ provided that $P_i$ and $P_j$ are the nearest neighbors in the grid. 

Matrices $\mathbf M$ and $\mathbf L$ depend only on the grid and are calculated accurately.
\[m_{ij}=S(O_{ij})(1+\delta_{ij})/12, \]
where $S(O_{ij})$ is the area of $O_{ij}$ and $\delta_{ij}$ is the Kronecker delta. The calculation of the gradient $\nabla\varphi_i$ is described in detail in Ref.~15. We only note here that since $\varphi_i$ is linear on each triangle, therefore, its gradient is constant on each triangle, and the matrix elements $l_{ij} $ are also calculated accurately. Magnetic energy dependent matrix elements $f_{ij}$ and $g_{ij}$ can be calculated using numerical integration over grid triangles included in $O_{ij}$.

\section{Numerical examples}

To demonstrate the capabilities of the finite element method as applied to the Brown equation, we present results of two simulations. In both simulations, the parameter values for Fe presented in Ref.~1 will be used. Fe possesses cubic magnetic anisotropy with internal magnetic energy per unit volume that can by expressed as
\begin{equation}
\label{ca}
V_a(x,y,z) = K_1[(xy)^2+(yz)^2+(zx)^2]+K_2(xyz)^2,
\end{equation}
where $K_1$, $K_2$ are the anisotropy constants and $x,y,z,\,x^2+y^2+z^2 = 1$, are the guided cosines of magnetic moment $\bm M$, which can be considered as well as the Cartesian coordinates of a point on the surface of a unit sphere. The function (\ref{ca}) defined on a sphere has 26 critical points: 6 minima, 8 maxima and 12 saddle points. The directions corresponding to these points will be used as the directions of the applied magnetic field in further simulation.

For cubic anisotropy (\ref{ca}), the dimensionless function $\tilde V$ can be written in spherical coordinates as
\begin{eqnarray}
\tilde V(\theta,\phi) &&= \epsilon_a\left[\cot^2\theta + (1 + \kappa\cos^2\theta)\cos^2\phi\sin^2\phi\right]\sin^4\theta \nonumber \\
&&-\epsilon_h[(h_x\cos\phi + h_y\sin\phi)\sin\theta +  h_z\cos\theta],
\end{eqnarray}
where $\kappa = K_2/K_1$, $h_x,h_y,h_z$ are the Cartesian components (guided cosines) of the unit vector $\bm h$, and
\[\epsilon_a = \frac{vK_1}{k_BT},\; \epsilon_h = \frac{vM_sH}{k_BT}.\]

According to Ref.~1, $K_1$ = 4.8$\times10^5$ erg/cm$^3$, $K_2$ = 0.5$\times10^5$ erg/cm$^3$, therefore $\kappa = 0.104$. The saturation magnetization per unit volume $M_s$ at $T$ = 20$^\circ$C is  equal 1714.0 emu/cm$^3$. For definiteness, we consider a Fe particle of a cubic shape with an edge size 24 nm, therefore, the particle volume is $v=24^3$ nm$^3$. Using the given parameter values, one can obtain $\epsilon_a = 164.023$ at $T = 20^\circ$C (293 K). For simulation of magnetization the value of $\epsilon _h = 4\epsilon_a$ was taken.

All calculations were performed on the grid described above for $n=81$, the grid has $N_p=65612$ nodes and $N_t=131220$ triangles. Formula for the angular part of the gradient on a unit sphere
\[\nabla=\bm e_\theta\frac{\partial}{\partial\theta} +\frac{\bm e_\phi}{\sin\theta}\frac{\partial}{\partial\phi},\]
where ${\bm e}_\theta, {\bm e}_\phi$ are the angular basic unit vectors of the spherical coordinate system, was used to calculate $\nabla\tilde V$. Since component expression for $\bm u$ in spherical coordinates is $\bm u = (0,0,1)$, the matrix elements $g_{ij}$ were calculated by the formula
\[g_{ij} = \int_{O_{ij}}{\nabla\varphi_i\cdot\left(-\frac{1}{\sin\theta}\frac{\partial \tilde V}{\partial\phi}\bm e_\theta + \frac{\partial \tilde V}{\partial\theta}\bm e_\phi\right)\varphi_j\,d\Omega}.\]
The integrals in $f_{ij}$ and $g_{ij} $ were estimated numerically for each triangle of the grid by dividing the triangle into 9 equal triangles and using the prismoidal formula.

The first simulation presented here is connected with the process of magnetization of a single-domain particle in a constant magnetic field. This problem was considered for various purposes in several papers \cite{17,18,19,20,21,22}, where FPE was solved using expansion in spherical harmonics followed by the Laplace transform and the method of matrix continued fractions. 

\begin{figure}[h]
\centering
\includegraphics{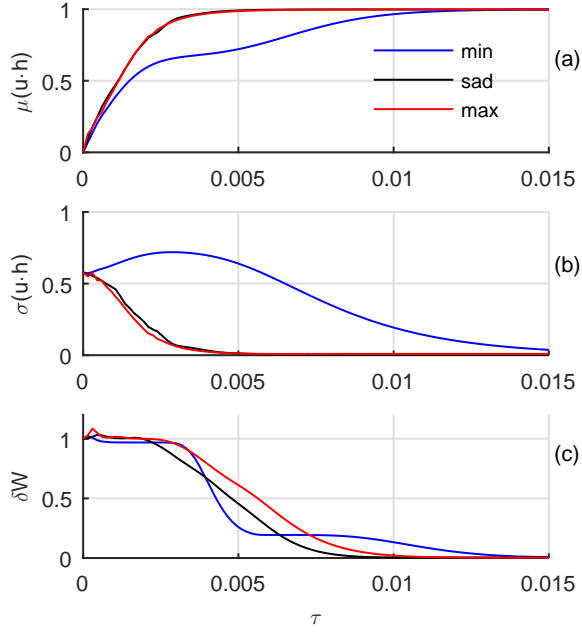}
\caption{Magnetization of a particle in constant field. (a) Average magnetization for three directions of applied field. (b) Mean square root deviation of average magnetization. (c) Normalized deviation of current distribution $W(\tau)$ from final distribution $W_H$. The value of damping parameter $\alpha=0.01$.}
\label{fig3}
\end{figure}

We present only the results of the FEM FPE solution without any physical interpretation. FPE was solved for the initial equilibrium distribution $W_0=Z^{-1}\exp(-vV_a/k_BT)$, $Z=\int{\exp(-vV_a/k_BT)\,d\Omega}$, for three directions of the applied field $\bm h$ corresponding to the critical points $V_a$. In Cartesian coordinates, these directions are expressed as ${\bm h}_{min} = (0,0,1)$, ${\bm h}_{max} = (1/\sqrt{3}, 1/\sqrt{3}, 1/\sqrt{3})$ and ${\bm h}_{sad} = (0,1/\sqrt{2}, 1/\sqrt{2})$. 

It is difficult to show the solution itself, that is, the probability density distribution over the unit sphere so that it is sufficiently informative. That is why we restrict ourselves to demonstration of the average magnetization $\mu(\bm u\cdot\bm h)$, -- the average projection of $\bm u$ on the direction of applied field $\bm h$:
\[\mu(\bm u\cdot\bm h) = \int{(\bm u\cdot\bm h)W(\bm u.\tau)\,d\Omega},\]
the mean square root deviation of the magnetization:
\[\sigma(\bm u\cdot\bm h) = \int{((\bm u\cdot\bm h)-\mu(\bm u\cdot\bm h))^2\,W(\bm u,\tau)\, d\Omega},\]
and the normalized deviation of current density distribution from final equilibrium distribution $W_H\propto \exp(-\tilde V)$:
\[\delta W=\frac{\int{|W(\tau)-W_H|\,d\Omega}}
{\int{|W_0-W_H|\,d\Omega}}.\]
All these integrals are calculated numerically using the numerical solution of FPE and the prismoidal formula. In particular,
\[\mu(\bm u\cdot\bm h)\approx \frac{1}{3}\sum_{k=1}^{N_t}{{((\bm u}_{k1}W_{k1}+ {\bm u}_{k2}W_{k2} + {\bm u}_{k3}W_{k3})\cdot\bm h)S_k}, \]
where ${\bm u}_{k1}, {\bm u}_{k2}, {\bm u}_{k3}$ are the radius-vectors of the vertices of $k$-th triangle, and $S_k$ is the area of $k$-th triangle.

Figure 3 shows three characteristics of magnetization process listed above obtained from numerical solution of FPE for initial equilibrium distribution at zero applied field. At turning on the external field the magnetic moment quickly alined with the direction of applied field. The mean square root of the average magnetization tends to zero, while the probability density distribution converges to the equilibrium (Boltzmann) distribution corresponding to the applied field $H$. 

\begin{figure}[h]
\centering
\includegraphics{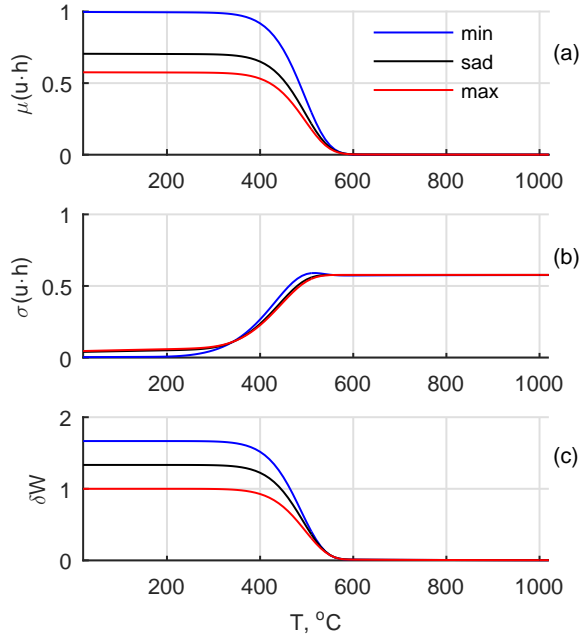}
\caption{Demagnetization of a particle under increasing temperature condition. The heating rate $k_T=10^{-3}$K. (a) Average magnetization. (b) The standard deviation of the magnetization. (c) The deviation of current solution from the Boltzmann distribution at current temperature.}
\label{fig4}
\end{figure}

FPE was resolved for the damping parameter $\alpha$ of 0.01. It should be noted that the stiffness of the system of equations strongly depends on $\alpha$ and rapidly increases with decreasing $\alpha$. The computation time also increases rapidly, since a very small time step is required to achieve the required accuracy. , The numerical examples presented here were calculated on a PC with a 4 GHz processor. With $\alpha = 1$, the calculation takes less than 1 minute, with $\alpha = 0.1$ it takes less than 5 minutes,
and 3-4 hours at $\alpha = 0.01$.

The second numerical example simulates the process of demagnetization that follows the moment of turning off the magnetic field considered in the first example. The process of demagnetization at constant (room) temperature is too long and cannot be adequately calculated from numerical solution of FPE because of accumulation of computational errors. To accelerate demagnetization we use heating of a particle with constant heating rate $k_T$, $T=T_0+k_T\tau$.

When $\epsilon_h = 0$ and the FPI numerical solution begins with the distribution $W_H$ -- the final distribution in the first simulation, the distribution $W(\tau)$ changes very quickly and if $\alpha <1$, it takes too much computational time to obtain the solution on PC. Therefore, we solved FPI at $\alpha=1$ with initial condition $W_0=W_H$ and the heating rate $k_T=10^{-3}$ K per unit of dimensionless time. The results, graphs of the average magnetization, the standard deviation of the magnetization, and the deviation of the current probability density distribution from the Boltzmann distribution at current $T$, in dependency on the particle temperature, are shown in Figure 4. Also, as in the first example, the graphs are plotted for the three directions of the external field at which the particle was magnetized. Here the deviation $\delta W$ is not normalized,
\[\delta W(\tau) = \int{|W(\tau)-W_B(T(\tau))|\,d\Omega}.\]

\begin{figure}[h]
\centering
\includegraphics{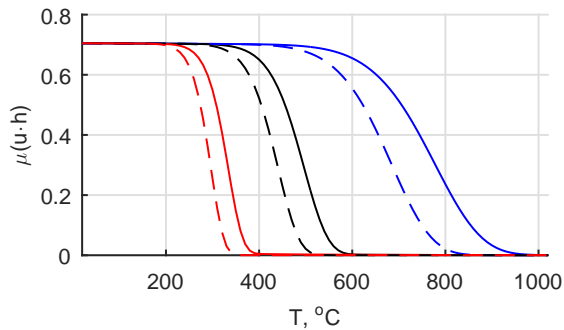}
\caption{Demagnetization of a particle under increasing temperature with different heating rates. Graphs of the average magnetization: blue line -- $k_T=10^{-1}$K, black line -- $k_T=10^{-3}$K, red line -- $k_T=10^{-5}$K. Solid lines -- $\alpha=1$, dashed lines -- $\alpha = 0.1$.}
\label{fig5}
\end{figure}

When heated, the average magnetization of the magnetized particle remains almost at the initial level until a certain temperature is reached, and then quickly drops. The temperature at which the falling magnetization becomes less than a certain value depends on the heating rate. For example, the average magnetization of a particle magnetized in an external field directed to the saddle point of $V_a$,
becomes less than 0.01 at $T_1=940^\circ$C if $k_t=10^{-1}$K, at $T_2=580^\circ$C if $k_t=10^{-3}$K, and at $T_3=390^\circ$C if $k_t=10^{-5}$K. These values are obtained from the FPE solution with $\alpha=1$. With a decrease in $\alpha$, these temperatures are also decrease. In particular, for $\alpha=0.1$ $T_1=820$K, $T_2=510$K, and $T_3=340$K.
As mentioned above, solving Eq.~(\ref{f1}) with the initial condition $W_H$ for $\alpha=0.1$ takes a lot of time on a PC. Therefore, here, as an initial condition, we take a solution to the equation with $\alpha = 1$ at the time $\tau = 1$. The graphs of the average magnetization versus temperature for three different heating rates are shown in Figure 5.

\section{Conclusion}

The results of this and previous works \cite{15} show that the finite element method allows one to efficiently solve Brown's FPI for single-domain particles with apparently much lower computational and programmatic efforts than the traditional method of decomposition into spherical harmonics. 

The procedure for generating triangular grids on the surface of a sphere using an inscribed icosahedron enables one to create an infinite sequence of regular (deterministic) grids with a fairly high quality. A comparison of the results obtained on grids with an increasing number of nodes makes it possible to draw a conclusion about the convergence of the numerical method. In this work, the calculations were performed on grids with $n$ = 72, 81, and 99. The results of calculations on these grids almost coincided, which makes it possible to conclude that it is inappropriate to increase the number of nodes to solve the problems considered.

All matrices of the system (\ref{f1}) have the same structure, nonzero elements in all matrices are in the same places. The number of nonzero elements, $N_{nz}=7N_p-12$,  is relatively small, so storing matrices does not require a lot of memory, and many calculations can be done on a regular PC. Another feature of the system (\ref{f1}) is that it is not resolved with respect to time derivatives. Therefore, for its numerical solution, it is necessary to use special codes for linear implicit systems, for example, the LSODIS codes. All the numerical examples presented in the paper were run on a regular PC using the MATLAB environment. To solve the system (\ref{f1}), the MATLAB function 'ode15s' was used. However, as parameter $\alpha$ decreases, the system stiffness increases rapidly, and solving a problem on a PC becomes problematic due to a very long computational time. Therefore, it is advisable to develop special codes for solving the system (\ref{f1}), taking into account its features and capabilities of multi-core processors.


\end{document}